# Solar Flare Detection Method using Rn-222 Radioactive Source

Jonathan Walg, Anatoly Rodnianski and Itzhak Orion

Nuclear Engineering, Ben Gurion University of the Negev, Beer Sheva, Israel

Abstract

Solar neutrino detection is known to be a very challenging task, due to the minuscule absorption cross-section and mass of the neutrino. One research showed that relative large solar-flares affected the decay-rates of Mn-54 in December 2006. Since most the radiation emitted during a solar flare are blocked before reaching the earth surface, it should be assumed that such decay-rate changes could be due to neutrino flux increase from the sun, in which only neutrinos can penetrate the radionuclide.

This study employs the Rn-222 radioactive source for the task of solar flare detection, based on the prediction that it will provide a stable gamma ray counting rate. In order to ascertain counting stability, three counting systems were constructed to track the count-rate changes.

The Rn-222 count-rate measurements showed several radiation counting dips, indicating that the radioactive nuclide can be affected by order of magnitude neutrino flux change from the sun. We conclude that using the cooled Radon source obtained the clearest responses, and therefore this is the preferable system for detecting neutrino emissions from a controlled source.

Keywords: gamma, neutrino, decay-rate, radiation, solar-activity

Introduction

Only one report has been published regarding the influence of solar flares on radioactive half-life [1]. This occurred in December 2006, when, for the first time, a study found that high-flux x-ray flares (class X – M) correlate to measured Mn-54 gamma radiation count-rate discrepancies. Mn-54 is an electron-capture radioactive nucleus that produces a gamma rays emitter, excited Cr-54, with a 312-day half-life [2]. The hypothesis that solar neutrino flux variations cause these count-rate discrepancies was presented by Jenkins and Fischbach [1]. Although the involvement of neutrinos is widely considered responsible for these decay rate variations, their part in radioactive decay is overlooked by nuclear physics models.

According to previous studies, measurements of half-life radioactive sources showed an annual periodical variation, despite the customary notion that radioactive decay should be considered a physical constant for each radionuclide. Alburger et al. (1986) conducted an experiment [3] in which decay rates of Si-32 and Cl-36 were simultaneously measured using the same detector system, and annual variations in count-rates were observed to differ in both amplitude and phase. Hence Alburger et al. concluded that half-life varies due to an annual periodical effect. Yet one recent publication by Sturrock, Steinitz and Fischbach [4] analyzes long-term (i.e., 10 years at 15-minute intervals) measurements of Rn-222 decay data using spectrograms of the measured gamma radiation followed by the Rn-222 alpha particle emission, suggesting that Rn-222 alpha particle emissions can present an annual periodical count-rate change. A publication by Pommé et al. [5], which includes data analysis of the Rn-222 annual periodical measurements, called into question the findings reported by Sturrock, Steinitz and Fischbach [4]. Indeed, Pommé et al. rule out annual variations at extremely sensitive levels (the solar neutrino flux varies by ±8% during the year due to the changing earth-sun distance). However, the current study aims to examine the detection of order of magnitude solar flare variations [5].

Solar x-ray flares occur when the sun's activity increases. It is evident that an 11-year sunspot cycle is related to solar activity, and therefore there is a greater probability of solar x-ray flares occurring in the higher solar activity phase of the cycle [6-7]. We are currently at the lowest phase of the solar activity cycle, and although the appearance of solar flares cannot be accurately predicted, maximal solar activity should occur during the years 2024–2025. The appearance of solar flares indicates neutrino flux 3-4 order of magnitudes increment, since in the flare process protons can be accelerated to energies in the range of GeV. The detailed solar flare neutrino production process and calculations were reported by Ryazhskaya et al. [8] that also predicted the possibility to detect these neutrino flux changes on Earth. The solar x-ray flare phenomenon is thought to be related to the particle transfer loop from the sun to the corona [9]; in addition, since they can interact with the earth's ionosphere, several satellites have been launched with the aim of measuring these flares and reporting their appearance time and magnitude. A series of GOES (Geostationary Operational Environmental Satellites) operated by the Space Weather Prediction Center, National Oceanic and Atmospheric Administration provides measured solar x-ray flux daily data, which is reported in units of $W/m^2$ for each minute. This x-ray flux is classified as A, B, C, M, or X according to peak flux magnitude, where class A, the lowest flux, is less than $10^{-7}$ $W/m^2$, X is above $10^{-4}$ $W/m^2$, and the difference from class to class is 10-fold.

Method

Three experimental setups of radiation measurement systems of NaI(Tl) detectors (2" diameter by 2" length) were installed in an underground laboratory, each facing a standard Ra-226 (100 kBq) source producing Rn-222. In the first setup, the Rn gas was run from the source Ra-226 chamber via a pipe toward the detection system with two NaI(Tl) detectors, as shown in Figure 1. The second system consisted of a Ra-226 (100 kBq) chamber that was pumped to the pressure of 1.33 Pa and sealed prior to measurements. One NaI(Tl) detector faced the gas Rn-222 in the chamber, as shown in Figure 2. A third counting system consisted of one NaI(Tl) detector facing an Rn-222 source in a chamber that was placed inside a freezer with a fixed temperature of -40° C in order to reduce the gaseous motion in the container and improve the counting stability. Each of the detectors in all three systems was shielded with a 5 cm thickness of lead. The lab walls and ceiling were made of 30 cm thick concrete, and the entire system was surrounded by 5 cm of lead.

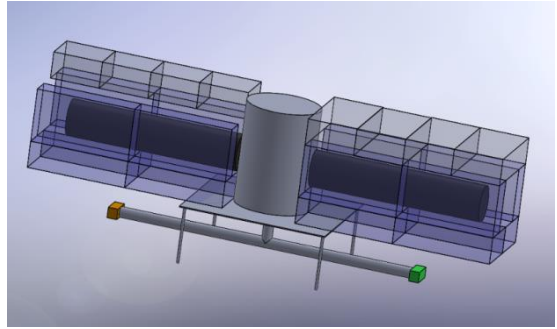

*Figure 1: Illustration of the first counting system: The Ra-226 source chamber emits Rn-222 gas into a pipe facing two NaI(Tl) detectors.*

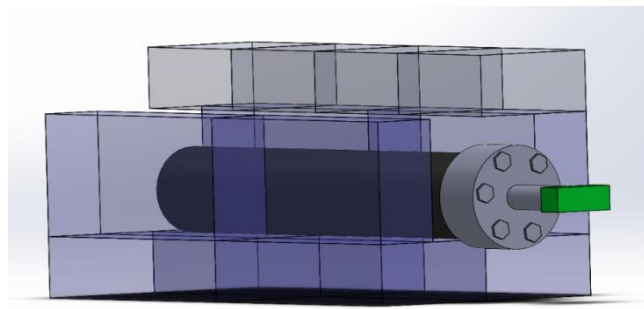

*Figure 2: Illustration of the second counting system: One NaI(Tl) detector faces the Ra-226 source emitting Rn-222 in a sealed chamber.*

All detectors in the three setups were connected to a data logger (DL), CR800, manufactured by Campbell Scientific, which remotely collected and submitted data to a computer that is remotely controlled for access to the DL and collected data. Every 15 minutes, gamma counts from each detector were integrated and tallied. The laboratory was permanently locked to avoid the influence of any other stimuli and unexpected radiation perturbations, and it was also environmentally controlled in terms of temperature and clean-air flow in order to reduce detector efficiency dependence.

Since oscillations in gas motion were detected in the first Rn-222 system, we revised the source conditions in the second system. Accordingly the Rn-222 was pumped from the Ra-226 source container and sealed before measurements in order to avoid the periodical puffing pattern. The Rn-222 gas built up an initial vacuum of 1.33 Pa during the first 12 days of measurements in the container.

The third system, as shown in Figure 3, consisted of one NaI(Tl) detector, with the Ra-226 radiation source inside a freezer set at a low temperature, minus forty centigrade (-40±1°C). The Rn-222 was transferred from the Ra-226 source container into a sealed cell. According to the laws of thermodynamics, gas vibrations should decline at relatively low temperatures (or low pressure).

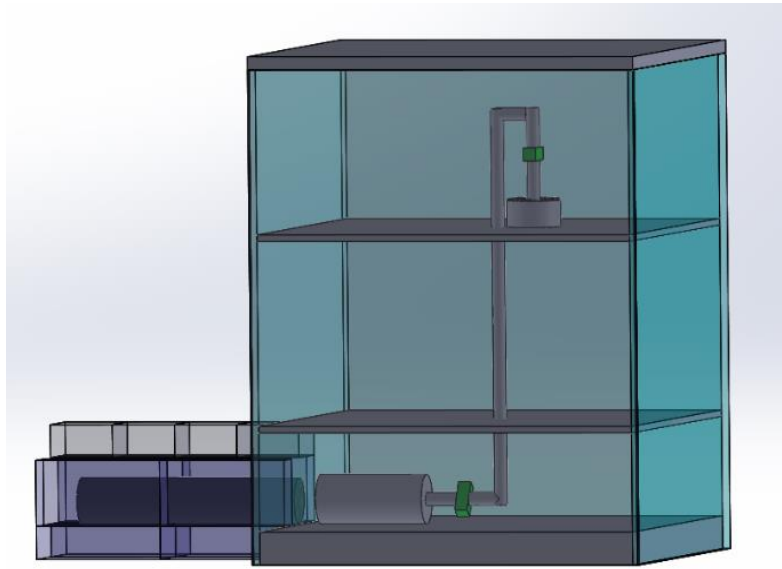

*Figure 3: Illustration of the third system: One NaI(Tl) detector and Ra-226 source emitting Rn-222 inside a freezer.*

Laboratory temperature, measured throughout the experiments, was maintained stable at 18°C (±1°). Temperature stability is required for such delicate changes since scintillation efficiency can be affected by temperature differences [10], even though peak counts are much more sensitive to these temperature changes compared to total-counts [11]. PM-11 background counts, measured for a two-day period, remained around the level of 700 cpm.

Solar flares were traced on a daily basis via the SpaceWeatherLive website [12], intended for uses related to astronomy, space, space-weather, aurora, etc. This website reports flare details and presents their graphs, categorizing them by flare intensity and time.

Results

The first system, with two NaI(Tl) detectors, operated continuously from August 22, 2018. On October 12-13, 2018, five solar flare events occurred. As a response to the solar flare events, few suspicious deep valleys were observed in both detectors.

Due to many fluctuations in the Rn-222 gas, changes were made to the configuration of the system. According to the laws of thermodynamics, when a gas is at low temperature or low pressure, there should be a reduction in its vibrations. Therefore, two new Rn-222 systems were initiated: the 'second system' – an Rn-222 source at low pressure, started during March 2019, and the 'third system' – an Rn-222 source at low temperature, initiated during May 2019.

The 'second system' responded to X-ray solar flares that occurred on March 22, 2019, as shown in Figure 4a. This was expressed by two dips in the detector's count-rate, as presented in Figure 4b. It was found that only relatively strong flares, such as class C, led to a substantial response in the 'second system'.

a

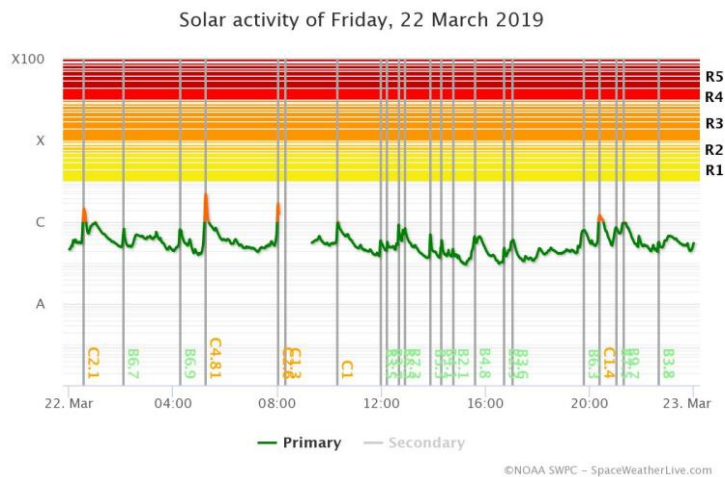

b

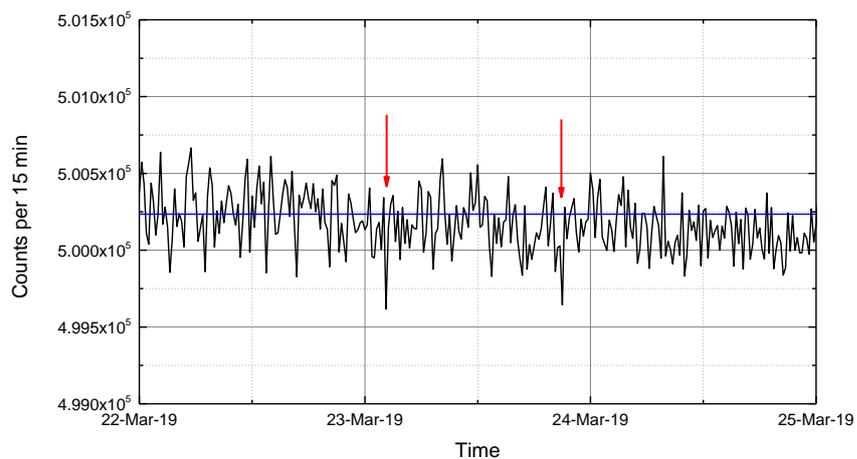

*Figure 4: a. Multiple solar flares that occurred on March 22, 2019. Full image from SpaceWeatherLive (with permission) [12]; b. Count-rate responses of the NaI(Tl) detector to solar flares: Red arrows indicate the revealed anomalies.*

The 'third system' was operated from May 2019. Between May 5 and 6, several X-ray solar flare events were recorded, as summarized in Table 1.

*Table 1: Several solar flare occurrence times (UTC+3) and their flux during May 2019. The flare flux in nW/m$^2$ at the beginning, end, and maximum are listed [12].*

| Date | start flare time UTC+3 | start flux (nW/m²) | max flare time UTC+3 | max flux (nW/m²) | end flare time UTC+3 | end flux (nW/m²) |
|---|---|---|---|---|---|---|
| 5.MAY.19 | 0:45 | 112 | 1:45 | 2120 | 3:00 | 110 |
| | 14:20 | 132 | 14:50 | 720 | 15:00 | 128 |
| | 15:30 | 119 | 16:40 | 980 | 17:10 | 118 |
| 6.MAY.19 | 0:00 | 175 | 0:45 | 610 | 4:10 | 138 |
| | 7:45 | 163 | 8:10 | 9970 | 8:35 | 156 |
| | 10:00 | 134 | 10:40 | 1700 | 11:30 | 280 |
| | 11:30 | 280 | 11:50 | 2000 | 12:55 | 206 |
| | 12:55 | 206 | 13:00 | 1500 | 13:50 | 150 |

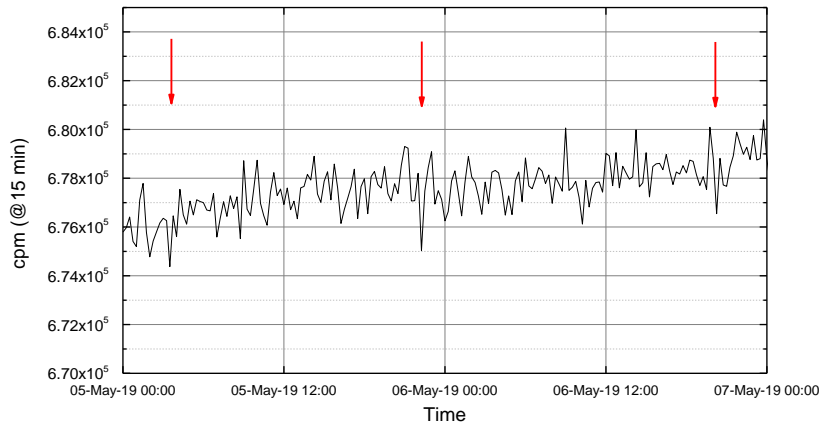

*Figure 5: The 'third system' count-rate NaI(T1) detector responses to solar flares: Red arrows indicate the revealed anomalies.*

Table 1 indicates that the x-ray solar flares events were discrete, with maximum flux varying from 610 to 9970 nW/m$^2$ (class B to class C). The detector response, as shown in Figure 5, recorded several dips due to x-ray solar flare events on May 5-6, 2019. The three peaks corresponded to the four groups of flares outlined in Table 1 (the 610 nW/m$^2$ is too low to be detected).

On May 15, 2019, a rapid and isolated class C solar flare event was registered at 22:25 (UTC+3) with a maximal flux of 1670 nW/m$^2$. The count-rate of the 'third system' clearly decreased in response to this event. In Figure 6, counts-per-hour are shown next to the event time, allowing us to present the statistical significance of the system detection. The average was 2.73866E+6 counts-per-hour, SE of mean of 283.5 counts-per-hour, and the response was 2.73788E+6 counts-per-hour (780 counts less), which are 2.75 times SE-of-mean, hence the count-rate tally exceeded the critical limit of detection.

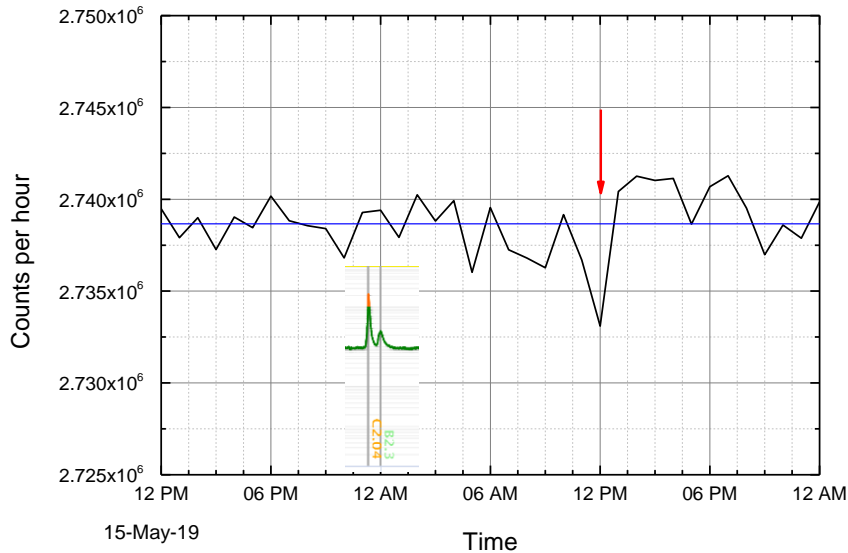

*Figure 6: The 'third system' counts-per-hour response to a rapid and isolated class C solar flare event on May 15, 2019 (presented next to the event time).*

The statistics significance of several results were calculated and presented in the Appendix of this paper.

Discussion

The results of the 'first system' implied that the Rn-222 system demonstrates a response to neutrinos, however it was difficult to isolate clear signals. Our findings encouraged us to develop improved, stabilized Rn-222 systems.

Throughout this study we measured counts for every 15 minutes. However, in the graphical presentation of the detector's response, we accumulated the counts into counts per hour data. This summation should be performed only when the system has reached a steady-state condition.

Therefore, the 'third system' was deemed preferable as a detector for the solar neutrino.

Conclusions

This study found that in the first Rn-222 system, the radioactive gas responds to solar flares within a few hours with a reduction in its count-rate (the half-life altered due to changes in neutrino flux from the sun). We constructed two new Rn-222 systems, one with a low vacuum and the other with a low temperature.

The second system also demonstrated a response to solar flares, yet sometimes encountered problems in detecting a response due to system fluctuations. The Rn-222 third system, similarly to the two previous systems, showed a response to solar flares. To reduce the noise, the count-rate was collected once every hour, instead of every 15 minutes.

Our measurements, showing at least six radiation counting dips, indicate that a radioactive nuclide can be affected by order of magnitude neutrino flux change from the sun. According to these results, the 'third system' is preferable for detecting artificial neutrino emission, and this requires further investigation.

The current nuclear models do not include a mechanism of neutrino absorption in an alpha emitter nucleus, while this reported phenomenon indicates that the neutrino could be interacting with an unstable nucleus, which might comprise unknown neutrino resonance absorption. Our findings, as reported in this work, encourage further theoretical studies regarding neutrino interaction with radioactive matter.

Appendix

In order to ascertain signal detectability for the measurements reported in this paper, we followed the method of limits-of-detectability as described by Knoll in the book "Radiation Detection and Measurement" (chapter 3 section VI)   [11]. The description by Knoll shows how to find real activity above background, however in our case we have to inspect count rate decrease (signal) below the mean value of the count rate. Following the method presented by Knoll we calculate for the dip the critical level ($L_C$) using the neighboring count rates moments, and comparing the counts at the dip to the critical level. $L_C$ is equal to 2.326 times standard deviation ("a 95% probability that a random sample will lie below the mean plus $1.645\sigma$"), therefore while the presented dip counts are below the $L_C$ value, then a reliable signal was detected. Selected treatments for the presented dips are shown as follows:

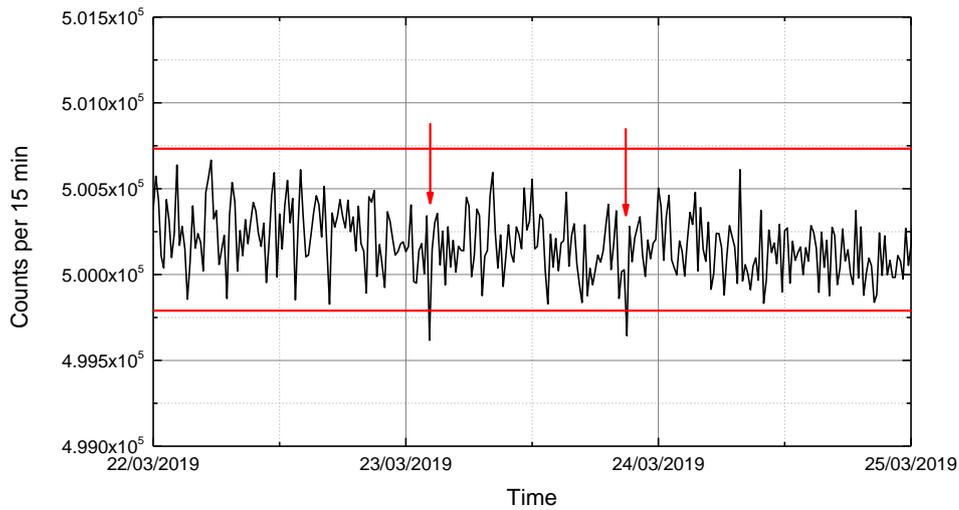

*Figure A1: same as Fig. 5b.: Count-rate responses of the NaI(T1) detector to solar flares: Red arrows indicate the revealed anomalies; 2.5σ and -2.5σ limit levels are presented.*

$Mean = 500261.435$

$\sigma = 188.6$

$L_C = 2.326 \cdot \sigma = 2.326 \cdot 188.6 = 438.68$

$Mean - L_C = 500261.435 - 438.68 = 499822.565$

$signal = 499616.8$

$diff = |signal - (Mean - L_C)| = 205.76$

In this dip 205.76 counts exceeded the critical level, which is even more than 1σ.

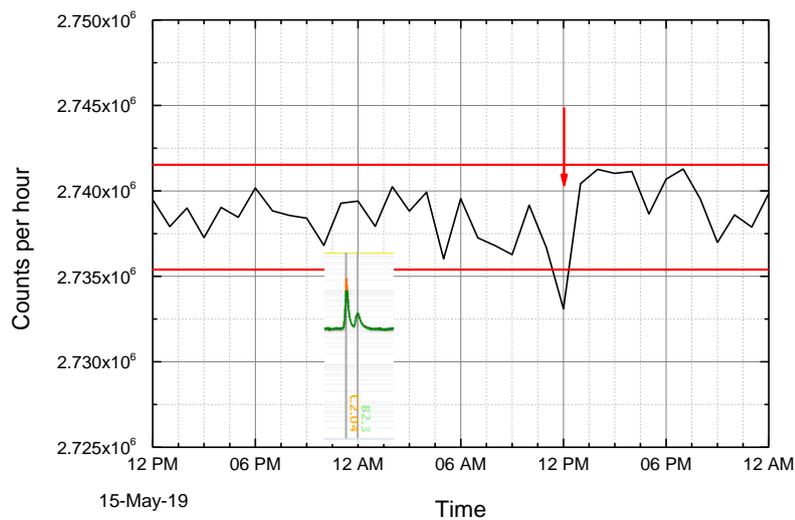

*Figure A2: same as Fig.7: The 'third system' counts-per-hour response to a rapid and isolated class C solar flare event on May 15, 2019. 2.5σ and -2.5σ limit levels are presented.*

$$Mean = 2.73846 \cdot 10^6$$
$$\sigma = 1227.118$$
$$L_C = 2.326 \cdot \sigma = 2.326 \cdot 1227.118 = 2896$$
$$Mean - L_C = 2.73846 \cdot 10^6 - 2896 = 2.735564 \cdot 10^6$$
$$signal = 2.733093 \cdot 10^6$$
$$diff = |signal - (Mean - L_C)| = 471$$

In this dip 471 counts exceeded the critical level.

These dip results are deemed statistically significant.

Note that the results from the first system were too fluctuating, therefore these findings were treated using signal processing only. Yet signal processing is a valid data analysis.